\documentclass[twocolumn]{aastex63}
\usepackage[]{graphicx}

\shorttitle{Formation of multiple planet systems}
\shortauthors{Y.-C. Lin et al.}

\begin{document}

\title{Formation of multiple-planet systems in resonant chains around M dwarfs}

\correspondingauthor{Yuji Matsumoto}
\email{ymatsumoto@asiaa.sinica.edu.tw}

\author{YU-CHIA LIN}
\affiliation{Department of Physics and Astronomy, University of New Mexico, Albuquerque, NM, USA}
\affiliation{Institute of Astronomy and Astrophysics, Academia Sinica, No.1, Sec. 4, Roosevelt Rd, Taipei
10617, Taiwan}
\affiliation{Department of Electrical Engineering, National Taiwan University, Taipei 10617, Taiwan}

\author[0000-0002-2383-1216]{Yuji Matsumoto}
\author[0000-0002-5067-4017]{Pin-Gao Gu}
\affiliation{Institute of Astronomy and Astrophysics, Academia Sinica, No.1, Sec. 4, Roosevelt Rd, Taipei
10617, Taiwan}%

\begin{abstract}

Recent observations have revealed the existence of multiple-planet systems composed of Earth-mass planets around late M~dwarfs.
Most of their orbits are close to commensurabilities, which suggests that planets were commonly trapped in resonant chains in their formation around low-mass stars.
We investigate the formation of multiple-planet systems in resonant chains around low-mass stars.
A time-evolution model of the multiple-planet formation via pebble accretion in the early phase of the disk evolution is constructed based on the formation model for the TRAPPIST-1 system by \cite{ormel2017formation}.
Our simulations show that knowing the protoplanet appearance timescale is important for determining the number of planets and their trapped resonances: as the protoplanet appearance timescale increases, fewer planets are formed, which are trapped in more widely separated resonances.
We find that there is a range of the protoplanet appearance timescale for forming the stable multiple-planet systems in resonant chains.
This range depends on the stellar mass and disk size.
We suggest that the protoplanet appearance timescale is a key parameter for studying the formation of multiple-planet systems with planets in resonant chains around low-mass stars.
The composition of the planets in our model is also discussed.

\end{abstract}


\section{Introduction} \label{sec:intro}

Recent observations have revealed the existence of multiple-planet systems around low-mass stars, such as TRAPPIST-1 \citep{gillon2017seven,luger2017seven}, YZ~Ceti \citep{Astudillo-Defru+2017b, Stock+2020}, Teegarden's star \citep{Zechmeister+2019}, and GJ 1061 \citep{Dreizler+2020}.
These planets have approximately one Earth mass and are orbiting around stars with masses of $\sim 0.1M_{\odot}$, where $M_{\odot}$ is the solar mass.
Except for those of the planets orbiting around Teegarden's star, their period ratios are close to the period commensurabilities.
Although more multiple-planet systems around low-mass stars are needed to perform statistical analyses, the fraction (3/4) of resonant or near resonant planetary systems around low-mass stars suggests that planets in or near resonant chains around low-mass stars would be more abundant than those around stars with masses of $\sim 1M_{\odot}$ \citep[][]{Fabrycky+2014,Winn&Fabrycky2015}.

\citet{ormel2017formation} proposed a scenario for the formation of compact planetary systems around TRAPPIST-1, which has about $0.08 M_{\odot}$.
This scenario includes the following stages:
a planetesimal forms around the H$_2$O iceline (hereafter iceline) owing to the streaming instability;
the planetesimal grows through pebble accretion to a protoplanet with the isolation mass while migrating inward;
the protoplanet stops migration around the magnetospheric cavity of the disk;
the next few protoplanets are formed sequentially in the similar manner and are trapped in resonant chains;
after the disk gas depletion, some mechanisms move planets out of resonance;
finally, planets reside at the current orbits.
This analytical model was further investigated by a numerical study \citep{Schoonenberg+2019}.
Furthermore, population synthesis simulations showed the mass distribution of protoplanets around low-mass stars based on the similar concept \citep{Liu_B+2019b, Liu_B+2020, Dash&Miguel2020}.

The aforementioned studies provide us insights into the accretion of protoplanets around M~dwarfs.
However, the formation of multiple planets in the chains of resonant orbits in the pebble-driven core accretion scenario has not been fully understood.
In particular, the configuration of planets in resonant chains would be affected by the planetesimal forming efficiency.
Recent studies showed that the streaming instability can be either promoted or hampered by turbulent stirring \citep{CL2020,Gole2020,Schafer+2020,Umurhan2020}.
The efficiency of the planetesimal formation through the streaming stability is also affected by the size distribution of dust \citep[e.g.,][]{Bai&Stone2010_apj722, Krapp+2019}.
These uncertainties may lead to a wide range of the planetesimal formation time.
The timescale of the subsequent growth of planetesimals to protoplanets depends on the uncertain size distribution of planetesimals formed via the streaming stability \cite[e.g.,][]{Liu_B+2019}.
It is expected that a longer appearance interval of a protoplanet that grows to be a planet leads to systems composed of a smaller number of planets in more widely separated resonances, which would not cause an orbital instability \citep{matsumoto2012orbital}.

In this study, we investigate the formation of multiple-protoplanet systems in which the protoplanets are trapped in resonant chains.
The time-sequential evolution of the protoplanets is considered based on the pebble-driven core accretion scenario \citep{ormel2017formation, Schoonenberg+2019}.
Moreover, the condition of the resonant trapping of protoplanets growing through pebble accretion is presented as a function of the protoplanet appearance time, stellar mass, and disk size.
The protoplanets trapped in resonant chains are referred to as ``planets" in this paper.
Our simulation results also show in which resonant chain the planets are trapped.

The remainder of this paper is organized as follows.
The model is described in Section \ref{sec:model}.
We present the results of our analytical simulations in Section \ref{sec:simulation}.
Section \ref{sec:discussion} discusses the influences of our assumptions on the results.
Finally, the conclusion is summarized in Section \ref{sec:conclusion}.

\section{Model} \label{sec:model}

\subsection{Outline}\label{subsec:scenarios}

We basically follow the pebble-driven core accretion scenario proposed in \cite{ormel2017formation}.
We consider a protoplanetary disk composed of gas and dust.
First, icy dust grains grow through coagulation.
As their masses increase, the grains become gradually decoupled from the gas according to their increasing Stokes number $\tau_{\rm p}$, which is the dimensionless stopping time.
We assume that the grains begin to drift inward when $\tau_{\rm p}=0.05$ \citep[e.g.,][]{birnstiel2012can, Okuzumi+2012}.
These grains are called ``pebbles" in this study.
Their growth timescale $t_{\rm peb}$ is a function of the dust-to-gas mass ratio \citep[][]{birnstiel2012can, Okuzumi+2012};
$t_{\rm peb} = \xi (Z_0 \Omega_{\rm Kep})^{-1}$ is the timescale over which the pebbles begin to drift inward, where the global dust-to-gas mass ratio $Z_0$ is 0.02, $\Omega_{\rm Kep}$ is the Kepler frequency, and $\xi=10$ \citep{ormel2017formation}.
The radial drift of pebbles ensues from the inside to outside of the disk.
The first icy pebbles form at $500/(2\pi) T_{\rm Kep,ice}\simeq 80 T_{\rm Kep,ice}$ and the final icy pebbles form at $t_{\rm peb, out}\simeq 80 T_{\rm Kep,disk}$, where $T_{\rm Kep}$ is the Kepler time; $T_{\rm Kep, ice}$ is $T_{\rm Kep}$ at the iceline and $T_{\rm Kep,disk}$ is $T_{\rm Kep}$ at the disk outermost radius ($R_{\rm disk}$).

The pebbles drift inward and create a pileup structure near the iceline \citep[e.g.,][]{schoonenberg2017planetesimal}, which is simply described by a Gaussian function of the pebble flux in our model.
The pebble flux in the region exterior to the iceline is derived from the growth and drift of pebbles.
When the dust density at the pileup peak exceeds the gas density, planetesimals are formed via the streaming instability.
An embryo is formed through the pebble accretion of scattered planetesimals and/or via the runaway growth of planetsimals \citep{Liu_B+2019,Schoonenberg+2019}.
We focus on the growth of the embryo, which is the precursor of a (proto)planet.
We use $t_{\rm pl}$ to express the appearance timescale of a 100~km sized embryo through the streaming instability and subsequent growth.
Hereafter, we call embryos as protoplanets for simplicity.
The protoplanet accretes the surrounding pebbles \citep[e.g.,][]{Ormel&Klahr2010, Lambrechts&Johansen2012}.
After the protoplanet has gained sufficient mass, it migrates inward owing to the protoplanet-disk tidal interactions \citep[e.g.,][]{Ward1986,Lin&Papaloizou1993}.
The mass growth of the protoplanet through pebble accretion lasts until its mass reaches the pebble isolation mass \citep{Lambrechts+2014}.
Afterward, the migration of the first protoplanet is stopped at the magnetospheric cavity of the disk.
The following protoplanets are sequentially trapped in mean motion resonances based on the competition between the resonant libration timescale and the migration timescale through the resonant width \citep{Ogihara_2013}.

We perform simulations of the protoplanets in this growth scenario by changing their appearance time ($t_{\rm pl}$), stellar mass ($M_\star$), and disk size ($R_{\rm disk}$).
Our simulations provide the number of planets trapped in resonant chains and the combinations of various resonant chains in a planetary system,
which affect their subsequent orbital evolutions (i.e., whether they cause orbital instabilities or not \citep{matsumoto2012orbital, Matsumoto&Ogihara2020}).
The details of the proposed model are presented in the following subsections.

\subsection{Disk} \label{subsec:disk}

The disks are divided into the two regions because the pebble formation timescales ($\simeq 80T_{\rm Kep}$) are shorter than the viscous timescales \citep{ormel2017formation}.
The inner disk, which is the region interior to the iceline, is viscously relaxed.
In this region, the gas surface density is given by $\Sigma_{\rm g,in} = {\dot M}_{\rm g}/3\pi\nu$ \citep{Lynden-Bell&Pringle1974}, where the viscosity is $\nu=\alpha (hr)^2 \Omega_{\rm Kep}$, $\alpha=10^{-3}$ in our model, $h$ is the disk aspect ratio, and $r$ is the orbital radius \citep{shakura1973black}.
The gas accretion rate depends on the stellar mass \citep{manara2015x}:
\begin{eqnarray}
	{\dot M}_{\rm g} &=& 10^{-10} M_\sun~{\rm yr^{-1}} \left(\frac{M_\star}{0.08M_\sun}\right)^{1.8}.
	\label{equ:gas_acc_mass}
\end{eqnarray}
Moreover, the disk aspect ratio is assumed constant; namely, $h=h_0=0.03$ \citep{ormel2017formation}.

In the disk outside of the iceline, the gas surface density is described by the power-law function:
\begin{equation}
	\label{equ:outer_gas_density}
	\Sigma_{\rm g} = \frac{M_{\rm disk}}{2\pi R_{\rm disk}^2}\left(\frac{r}{R_{\rm disk}}\right)^{-1},
\end{equation}
where the total disk mass is $M_{\rm disk}=0.04M_\star$ \citep{ormel2017formation}.
The disk aspect ratio is modeled by the following power-law function:
\begin{equation}
	\label{equ:out_h}
	h = 0.03 \left(\frac{r}{r_{\rm ice}}\right)^{1/4}.
\end{equation}
This function provides the continuous disk aspect ratios between the inner and outer disks and can be derived from the disk temperature $T\propto r^{-1/2}$ in the outer disk.

The gas temperature is expressed as follows:
\begin{eqnarray}
	\label{equ:T}
	T = 180\mbox{~K}\left(\frac{M_\star}{0.08M_{\odot}}\right)
	\left(\frac{h}{0.03}\right)^{2}
	\left(\frac{r}{0.1~{\rm au}}\right)^{-1},
\end{eqnarray}
for both the inner and outer disk parts.
The iceline is set at the disk location where the temperature is 170~K \citep{Hayashi1981}.
In our model, the iceline location ($r_{\rm ice}$) is proportional to $M_\star$.
Because the inner disk is considered viscously relaxed, our simulations start from the viscous timescale at the iceline:
\begin{eqnarray}
	\label{equ:viscous_time_ice}
	t_{\rm v, ice} &=&\frac{r_{\rm ice}^2}{\nu}
	\simeq 1.8\times 10^5 T_{\rm Kep,ice} \left( \frac{\alpha}{10^{-3}} \right)^{-1} \left( \frac{h}{0.03} \right)^{-2}
	\nonumber \\
	&\simeq& 2.2\times 10^4\mbox{~yr} \left(\frac{M_\star}{0.08M_{\odot}}\right) \left( \frac{\alpha}{10^{-3}} \right)^{-1} \left( \frac{h}{0.03} \right)
	.\nonumber\\
\end{eqnarray}
Thus, our model focuses on the early history of the formation of planetary systems.

The inner boundary of the disk coincides with the magnetospheric cavity radius:
\begin{eqnarray}
	\label{equ:r_c}
	r_{\rm c}
	&=& \zeta \left(\frac{B_\star^4 R_\star^{12}}{{\rm G}M_\star \dot{M}_{\rm g}^2}\right)^{1/7},
\end{eqnarray}
where $B_\star$ is the magnetic field strength of the stellar surface, $R_\star$ is the stellar radius, and $\zeta = (1/8)^{1/7}\simeq 0.74$ is a dimensionless factor ranging from 0.5 to 1 for an aligned dipole \citep[e.g.,][]{frank2002accretion, Chang_SH+2010}.
We adopt $B_\star=180$~G, and $R_{\star}=0.5R_{\odot}$ for $0.08M_{\odot}$ stars, where $R_{\odot}$ is the Solar radius \citep{Reiners+2009}.
We assume that the stellar radius increases linearly with increasing stellar mass.
This dependence is motivated by the estimation in \cite{Hayashi1966} and the observed empirical mass-radius relationship of M~dwarfs \citep{bayless20062mass}.
However, recent studies demonstrated that the stellar size is more complicated and the stellar mass dependence changes in the pre-main-sequence evolution when the inefficient injection of accretion heat is considered \citep[e.g.,][]{Kunitomo+2017}.
Ignoring this complication, we substitute the aforementioned values and Equation~(\ref{equ:gas_acc_mass}) into Equation~(\ref{equ:r_c}), and we obtain
\begin{eqnarray}
	\label{equ:r_c_mass}
	r_{\rm c}
	&\simeq& 0.92 \times 10^{-2} \mbox{~au}
	\left(\frac{B_\star}{180~\mbox{G}} \right)^{4/7} \left( \frac{M_\star}{0.08M_{\odot}} \right)^{7.4/7},
	\nonumber \\
\end{eqnarray}
which is approximately proportional to the stellar mass.

\subsection{Pebble} \label{subsec:pebble}

\subsubsection{Pebble Flux} \label{subsubsec:pebble_flux}
In our model, there is a pebble forming front at a certain orbital radius (Section \ref{subsec:scenarios}).
The orbital radius of the pebble front for $t=t_{\rm peb}$ is given by
\begin{equation}
	\label{equ:r_g}
	r_{\rm g} = \left(\frac{GM_\star Z_0^2 t^2}{\xi^2}\right)^{1/3}.
\end{equation}
After the pebbles have formed at $r_{\rm g}$, they drift inward with the velocity \citep{Adachi+1976, weidenschilling1977aerodynamics}
\begin{equation}
	\label{equ:v_r}
		v_{r} (\tau_{\rm p}) = -2 \eta \left(\frac{\tau_{\rm p}}{1+\tau_{\rm p}^2}\right) v_{\rm Kep},
\end{equation}
where the dimensionless pressure gradient is $\eta = 5h^2/4$ and $v_{\rm Kep}$ is the Kepler velocity in the outer disk.
With Equation (\ref{equ:out_h}), $v_{r}$ becomes
\begin{equation}
	v_{r} = - \frac{5}{2}h_0^2\left(\frac{\tau_{\rm p}}{1+\tau_{\rm p}^2}\right) v_{\rm Kep,ice},
\end{equation}
where $v_{\rm Kep,ice}$ is the Kepler velocity at the iceline.
Because $v_{r}$ does not depend on $r$ in the above equation, the pebble travel time from $r_{\rm g}$ to $r_{\rm ice}$ can be simply expressed as follows:
\begin{eqnarray}
	t_{\rm drift} &=& \frac{r_{\rm g}-r_{\rm ice}}{|v_{r}|} 
	.
\end{eqnarray}
When $t=t_{\rm peb}$, pebbles form and begin to drift inward at $r_{\rm g}$; their accretion rates are ${\dot M}_{\rm p}(r_{\rm g}, t_{\rm peb}) =2\pi r_{\rm g} {\dot r}_{\rm g}Z_0 \Sigma_{\rm g}$.
These pebbles reach the iceline when $t=t_{\rm peb}+t_{\rm drift}$.
The pebble mass flux at the iceline is given by
\begin{eqnarray}
	{\dot M}_{\rm p} (r_{\rm ice}, t) = {\dot M}_{\rm p}(r_{\rm g}, t_{\rm peb}) \frac{ dt_{\rm peb} }{ dt },
\end{eqnarray}
where $dt_{\rm peb}/dt$ is the correction due to $t_{\rm drift}$.
Owing to $t_{\rm drift}\propto r_{\rm g}\propto t_{\rm peb}^{2/3}$, the following expression holds:
\begin{eqnarray}
	\frac{ dt }{ dt_{\rm peb} } = 1 + \frac{2}{3} \frac{ t_{\rm drift} }{ t_{\rm peb} }.
\end{eqnarray}
Because $t_{\rm drift}/t_{\rm peb}\propto r_{\rm g}^{-1/2}$, $t_{\rm drift}/t_{\rm peb}$ decreases over time.
The pebble-to-gas mass flux ratio is as follows:
\begin{eqnarray}
	\label{equ:F_pg_nobump}
	\mathcal{F}_{\rm p/g}
	&=& \frac{\dot{M}_{\rm p}}{\dot{M}_{\rm g}}
	\nonumber\\
	&=&
	\frac{2}{3}
	\left( \frac{(2\pi)^2 Z_0^{5}}{\xi^{2}} \right)^{1/3}
	\frac{M_{\rm disk}}{\dot{M}_{\rm g} T_{\rm Kep,disk}^{2/3} t_{\rm peb}^{1/3} }
	\nonumber \\ &&\times
	\left( 1+ \frac{2}{3} \frac{t_{\rm drift}}{t_{\rm peb}} \right)^{-1}
	.
\end{eqnarray}
In addition, the pebble drift time is added as a correction term for the final pebble reach time:
\begin{eqnarray}
	\label{equ:t_end_c}
	t_{\rm end} &=& t_{\rm peb}(R_{\rm disk}) + t_{\rm drift}(R_{\rm disk})
	\nonumber\\
	&\simeq&
	\left[
	\frac{\xi}{2\pi Z_0}+
	\frac{1}{5\pi} h_0^{-2} \left(\frac{1+\tau_{\rm p}^2}{\tau_{\rm p}}\right) \left( \frac{r_{\rm ice}}{R_{\rm disk}} \right)^{1/2}
	\right]T_{\rm Kep, disk}
	\nonumber\\
	&\simeq&
	2.8\times10^5 \mbox{~yr}
	\nonumber \\ &&\times
	\left[
		1
		+
		0.58
		\left( \frac{h_0}{0.03} \right)^{-1}
		\left(\frac{\tau_{\rm p}}{0.05} \right)^{-1}
		\left( \frac{R_{\rm disk}}{100~\mbox{au}} \right)^{-1/2}
		\right. \nonumber \\ && \left.\times
		\left( \frac{M_\star}{0.08M_{\odot}} \right)^{1/2}
		\left(\frac{\xi}{10}\right)^{-1}
		\left( \frac{Z_0}{0.02} \right)
	\right]
	\nonumber \\ &&\times
	\left(\frac{\xi}{10}\right)
	\left( \frac{Z_0}{0.02} \right)^{-1}
	\left( \frac{M_\star}{0.08M_{\odot}} \right)^{-1/2}
	\left( \frac{R_{\rm disk}}{100~\mbox{au}} \right)^{3/2}
	\nonumber\\
\end{eqnarray}
The contribution of $t_{\rm drift}(R_{\rm disk})$ is 0.58 times of that of $t_{\rm peb}(R_{\rm disk})$ when $R_{\rm disk}=100$~au and $M_{\star}=0.08M_{\odot}$, and this ratio increases as the stellar mass increases.
When $t=t_{\rm end}$, we assume that the pebble flux is equal to zero at the iceline.
We note that our $\mathcal{F}_{\rm p/g}$ and $t_{\rm end}$ are different from those in \cite{ormel2017formation} since we consider the effect of the pebble drift.
This makes $\mathcal{F}_{\rm p/g}$ smaller and $t_{\rm end}$ longer compared to those in \cite{ormel2017formation}.
It is worth noting that $t_{\rm drift}/t_{\rm peb}\propto r_{\rm g}^{-1/2}$, which indicates that $t_{\rm drift}/t_{\rm peb}$ is larger than 0.58 at an early time.

The pebble flux inside the iceline is calculated in a similar way.
However, $\tau_{\rm s}=10^{-3}$ is used for silicate grains.
The Stokes number of the pebbles affects the drift timescale of a pebble going from $r_{\rm g}$ to $r$ when $r<r_{\rm ice}$; i.e.,
\begin{eqnarray}
	t_{\rm drift}(r) &=& \frac{r_{\rm g}-r_{\rm ice}}{|v_{r}(\tau_{\rm p})|} + \frac{r_{\rm ice}-r}{|v_{r} (\tau_{\rm s})|}.
\end{eqnarray}
Thus, the pebble-to-gas mass flux ratio inside the iceline is computed based on this drift time.

\subsubsection{Pebble-to-gas density ratio}

\begin{figure}
	\plotone{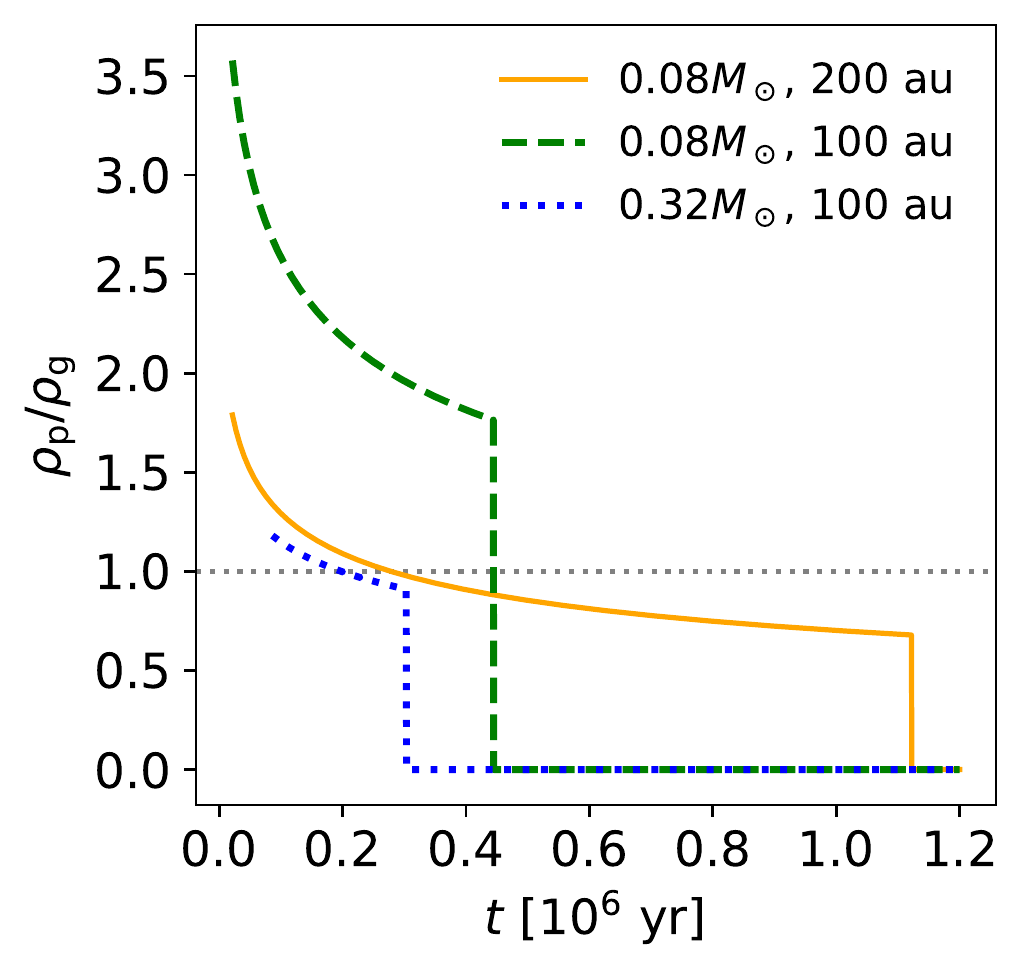}
	\caption{
		Time evolution of the pebble-to-gas density ratio at the iceline ($\rho_{\rm p}/\rho_{\rm g}$) is shown.
		Different curves show the results for the different parameters ($M_\star$, $R_{\rm disk}$).
		When $t$ exceeds $t_{\rm end}$ (Equation (\ref{equ:t_end_c})), $\rho_{\rm p}/\rho_{\rm g}=0$ because no pebbles are left in the outside of the iceline.
		The gray dotted line indicates $\rho_{\rm p}/\rho_{\rm g}=1$, which is the planetesimal formation criterion.
		}\label{fig:density_evolution}
\end{figure}

The pebble density at the midplane is important in the formation of planetesimals through the streaming instability \citep[e.g.,][]{youdin2005streaming, Johansen+2007} and the growth of planetesimals and protoplanets through pebble accretion \citep[e.g.,][]{Lambrechts&Johansen2012}.
Making use of the pebble surface density $\Sigma_{\rm p}={\dot M}_{\rm p}/(2\pi r v_r)$ and scale height $H_{\rm p}=\sqrt{\alpha/\tau_{\rm p}} H_{\rm g}$ \citep{dubrulle1995dust,Youdin&Lithwick2007}, the pebble-to-gas density ratio at the midplane is given by
\begin{eqnarray}
	\label{equ:p_gas_ratio}
	\frac{\rho_{\rm p}}{\rho_{\rm g}}
	&=& \frac{3}{5}\sqrt{\frac{\alpha}{\tau_{\rm p}}} f_{\rm ice} \mathcal{F}_{\rm p/g},
\end{eqnarray}
where $f_{\rm ice}$ represents the pileup profile around the iceline.
We adopt a Gaussian function for $f_{\rm ice}$ as follows:
\begin{eqnarray}
	f_{\rm ice} = \max\left( f_{\rm peak}\exp{\left( -\frac{(r-r_{\rm ice})^2}{2(r_{\rm ice} \delta_{\rm ice})^2} \right)} ,1\right),
\end{eqnarray}
where $\delta_{\rm ice}=0.05$ is the typical width of the iceline and $f_{\rm peak}=19$.
Our choice of $f_{\rm peak}$ is similar to $f_{\rm \Sigma, peak}$ in \cite{schoonenberg2017planetesimal}.
Although $f_{\rm peak}$ changes according to parameters such as $\alpha$, $\mathcal{F}_{\rm p/g}$ and ${\dot M}_{\rm g}$ \citep[e.g.,][]{schoonenberg2017planetesimal, hyodo2019formation}, in this study, a constant $f_{\rm peak}$ value is used in the simulations for simplicity.
We choose $f_{\rm peak}=19$ to produce a similar peak value of the density ratio to that in \cite{ormel2017formation} (Fig. 3) in the $\alpha=10^{-3}$ case.
Our $f_{\rm peak}$ value is, however, larger than \cite{ormel2017formation} since our $\mathcal{F}_{\rm p/g}$ is smaller (Equation (\ref{equ:F_pg_nobump})).
A larger $f_{\rm peak}$ value would be explained back-reaction of silicate grains \citep{hyodo2019formation}.

Figure \ref{fig:density_evolution} shows the time evolution of $\rho_{\rm p}/\rho_{\rm g}$ at the iceline for three cases.
The pebble-to-gas density ratio at the iceline decreases with time because the pebble mass flux decreases with time (Equation (\ref{equ:F_pg_nobump})).
When the disk size is small, $\rho_{\rm p}/\rho_{\rm g}$ is high but decreases quickly to zero.
In small disks, the pebble mass flux is high because the total disk gas mass ($M_{\rm disk}$) and metallicity ($Z_0$) are fixed and $t_{\rm peb}$ is short.
Around the massive stars, the pebble-to-gas mass flux ratio $\mathcal{F}_{\rm p/g}$ is low due to its inverse dependence on ${\dot M}_{\rm g}$ (Equation (\ref{equ:F_pg_nobump})) and a small value of $t_{\rm peb}$.

\subsection{Growth and Migration of Protoplanets} \label{subsec:planet}

\subsubsection{Growth of protoplanets}

We assume that a protoplanet forms through the streaming instability and subsequent growth at the iceline.
Given its radius is $\sim 100$ km, the protoplanet is assumed to have the initial mass of $4.2\times 10^{21}$~g.
The pebbles around the iceline pileup where $r>r_{\rm ice}(1-\delta_{\rm ice})$ would be a mixture of ice and silicates.
We assume that only icy pebbles exist in this region. Thus, the Stokes number of a pebble is $\tau_{\rm p}$.
In this region, the scale height of the icy pebbles is small.
The pebbles are concentrated around the disk midplane, and the pebble accretion takes place in a 2D manner (i.e., 2D mode).
In the region where $r<r_{\rm ice}(1-\delta_{\rm ice})$, there are silicate grains without the icy mantle, which are relatively small in size and thus are well coupled with gas.
Consequently, the scale height of silicate grains is almost equal to the gas scale height.
The silicate pebbles are widely distributed in the vertical direction and only those around the midplane are able to accrete onto protoplanets in a 3D manner (i.e., 3D mode).
In these two pebble accretion modes, the pebble accretion efficiencies are given by the fraction of the pebble accretion rate to the pebble flux \citep{Guillot+2014,ida2016radial, Ormel2017}:
\begin{eqnarray}
	\label{equ:eps_2}
	\epsilon_{\rm 2D}
	&\sim&
	0.1 \left( \frac{q_{\rm pl}}{10^{-5}} \right)^{2/3} \left( \frac{\tau_{\rm p}}{0.05} \right)^{-1/3} \left( \frac{h}{0.03} \right)^{-2},
	\\
	\label{equ:eps_3}
	\epsilon_{\rm 3D}
	&\sim&
	0.07 \left( \frac{q_{\rm pl}}{10^{-5}} \right) \left( \frac{h}{0.03} \right)^{-3},
\end{eqnarray}
where $q_{\rm pl}=M_{\rm pl}/M_{\star}$ is the protoplanet mass-to-central star mass fraction.
The pebble accretion efficiencies are less than 20\% even for a planet with the pebble isolation mass, and this is the reason why the effect of the pebble loss on the pebble density evolution caused by accretion is not considered in this study \citep[e.g.,][]{Lambrechts&Johansen2014,Guillot+2014}.
The pebble accretion rate is given by
\begin{eqnarray}
	{\dot M}_{\rm pl}= \epsilon f_{\rm ice} \mathcal{F}_{\rm p/g}\dot{M}_{\rm g}.
\end{eqnarray}
The protoplanets stop growing when the pebble isolation mass is reached \citep{Lambrechts+2014}:
\begin{eqnarray}
	M_{\rm iso} = h^3 M_{\star}.
	\label{equ:Miso}
\end{eqnarray}

\subsubsection{Planetary migration}\label{sec:migration}

As a result of planet-disk tidal interactions, protoplanets migrate inward based on either type I or type II migration depending on the gap opening in the gaseous disk.
These migration timescales are
\begin{eqnarray}
	t_{\rm I}
	&=& \frac{M_{\star} h^2 }{\gamma_{\rm I} q_{\rm pl} \Sigma_{\rm g} r^2 \Omega_{\rm Kep} }
	\nonumber \\
	&\simeq& 
	1.5\times 10^5\mbox{~yr}
	\left( \frac{q_{\rm pl}}{10^{-5}} \right)^{-1}
	\left( \frac{\gamma_{\rm I}}{4} \right)^{-1}
	\left( \frac{\alpha}{10^{-3}} \right)
	\left( \frac{h}{0.03} \right)^{4}
	\nonumber \\&&\times
	\left( \frac{M_{\star}}{0.08M_{\odot}} \right)^{-0.8},
	\label{equ:t_I}
	\\
	t_{\rm II} &\simeq& (1+0.04K)t_{\rm I} ,
	\label{equ:t_II}
\end{eqnarray}
where the factor $\gamma_{\rm I}$ depends on the disk temperature and surface density structures \citep{tanaka2002three,kley2012planet}, and the factor $1+0.04K$ is related to the gap around a protoplanet \citep{Kanagawa+2018},
\begin{eqnarray}
	K=q_{\rm pl}^2h^{-5}\alpha^{-1}
	= 4.1\left( \frac{q_{\rm pl}}{10^{-5}}\right)^2 \left( \frac{h}{0.03} \right)^{-5} \left( \frac{\alpha}{10^{-3}} \right)^{-1}.
	\nonumber \\
\end{eqnarray}
When the mass of a protoplanet approximately reaches $M_{\rm iso}$, a partial gap is opened, and the migration becomes slower than the type-I migration.
Applying $q_{\rm pl}=h^3$ to the above equation, the maximum value of $K$ is about 30, and thus the type-II migration timescale, $t_{\rm II}$, of the protoplanet becomes about twice as slow as $t_{\rm I}$.
It would be worth noting that the above migration timescales are derived in the absence of dust feedback, which would otherwise affect the migration timescales \citep{Kanagawa2019, Hsieh_HF&Lin_MK2020}.

By considering $\delta_{\rm ice} t_{\rm I} \approx t_{\rm grow} = M_{\rm pl}/{\dot M}_{\rm pl}$, where
\begin{eqnarray}
	\label{equ:t_gr}
	t_{\rm grow}
	&=&
	8\times 10^3 \mbox{~yr}
	\left( \frac{1 }{ \epsilon f_{\rm ice} \mathcal{F}_{\rm p/g} } \right)
	\left(\frac{q_{\rm pl}}{10^{-5} }\right)
	\left(\frac{M_\star}{0.08M_{\odot}}\right)^{-0.8}
	\nonumber\\
	&\simeq&
	4.2
	\times 10^3 \mbox{~yr}
	\left( \frac{\mathcal{F}_{\rm p/g}}{1} \right)^{-1}
	\left( \frac{f_{\rm ice}}{19} \right)^{-1}
	\left(\frac{q_{\rm pl}}{10^{-5} }\right)^{1/3}
	\nonumber \\ &&\times
	\left( \frac{\tau_{\rm p}}{0.05} \right)^{1/3}
	\left( \frac{h}{0.03} \right)^{2}
	\left(\frac{M_\star}{0.08M_{\odot}}\right)^{-0.8}
	,
\end{eqnarray}
a protoplanet crosses the iceline inner edge when its mass reaches
\begin{eqnarray}
	\label{equ:M_cr}
	M_{\rm cross} &\approx&
	0.98\times 10^{-5} M_\star\sqrt{ \mathcal{F}_{\rm p/g} \epsilon_{\rm 2D} f_{\rm ice} }
	\nonumber \\ && \times
	\left( \frac{\gamma_{\rm I}}{4} \right)^{-1/2}
	\left( \frac{\alpha}{10^{-3}} \right)^{1/2}
	\left( \frac{\delta_{\rm ice}}{0.05} \right)^{1/2}
	\left( \frac{h}{0.03} \right)^{2}
	.\nonumber \\
\end{eqnarray}
The fraction of the crossing mass ($M_{\rm cross}$) to the isolation mass is
\begin{eqnarray}
	\label{equ:M_cr_Miso}
	\frac{M_{\rm cross}}{M_{\rm iso}}
	&\approx&
	0.58
	\left( \frac{\mathcal{F}_{\rm p/g}}{1} \right)^{3/4}
	\left( \frac{f_{\rm ice}}{19} \right)^{3/4}
	\left( \frac{\tau_{\rm p}}{0.05} \right)^{-1/4}
	\left( \frac{\gamma_{\rm I}}{4} \right)^{-3/4}
	\nonumber \\ && \times
	\left( \frac{\alpha}{10^{-3}} \right)^{3/4}
	\left( \frac{\delta_{\rm ice}}{0.05} \right)^{3/4}
	\left( \frac{h}{0.03} \right)^{-3/2}.
\end{eqnarray}
This fraction reflects the composition of the protoplanets:
the early formed protoplanets are almost entirely composed of iceline materials, and the later protoplanets have less iceline materials.
Considering $\rho_{\rm p}/\rho_{\rm g}\geq 1$ at the iceline, $\mathcal{F}_{\rm p/g}\geq 0.62$ (Equation (\ref{equ:p_gas_ratio})), and $M_{\rm cross}/M_{\rm iso}\gtrsim 0.4$.
The fraction of iceline materials in our estimation is higher than that in \cite{ormel2017formation} due to the larger $f_{\rm ice}$.

After the protoplanets cross the iceline inner edge, a protoplanet is added within the iceline at every $t_{\rm pl}$.
Hence, the true time span between the appearance of protoplanets is given by the timescale on which a protoplanet leaves from the iceline region plus the appearance timescale of the next protoplanet, which includes the formation and growth of an embryo:
\begin{eqnarray}
	t &=& t_{\rm cross} + t_{\rm pl}
	\label{equ:t_c_tpl}
\end{eqnarray}
where $t_{\rm cross}$ is the timescale over which a protoplanet crosses the iceline and is given by $t_{\rm cross}\approx \delta_{\rm ice} t_{\rm I} $ where $q_{\rm pl}\approx M_{\rm cross}/M_\star$.
We note that the contribution of $t_{\rm grow}$ is included in $M_{\rm cross}$.

Because the migration timescales do not depend on $r$ (Equations (\ref{equ:t_I}) and (\ref{equ:t_II})), the protoplanets migrate inward exponentially over time.
After the first planet reaches $r_{\rm c}$, the other planets are trapped in mean motion resonances.
The resonance in which planets become trapped is determined by the comparison between the migration timescale and the critical migration timescale \citep{Ogihara_2013}.
Planets around low-mass stars tend to be trapped in the closest first-order resonances of the resonances that exist in the inner orbits than in the orbits of planets before the resonant trapping.
This is because the migration timescale increases and the critical migration decreases at $r_{\rm c}$ with decreasing $M_{\star}$; this allows planets to migrate more slowly and to be trapped in any first-order resonances.
As the orbital resonances narrower than 6:5 could be observationally rare \citep{Fabrycky+2014}, we restrict ourselves to the resonances from 2:1 to 6:5 for the resonant trapping in this study.

We consider that the innermost planets are at $r_{\rm c}$, no matter how many planets are trapped in resonances, for simplicity.
The location of the innermost planet would be determined by the torques on it.
The innermost planet does not only feel the inward migration (negative) torques of outer planets but also feel opposite direction (positive) torques due to the disk discontinuity \citep{Ogihara+2010, Liu_B+2017}.
The innermost planet stays at $r_{\rm c}$ when the positive torques are balanced with the migration torques.
Furthermore, planets in resonances do not cause orbital instabilities since their eccentricities are quickly damped by the tidal interaction with the disk gas \citep{Iwasaki+2001, Tanaka&Ward2004}.

\subsection{Termination of Planetesimal Formation}\label{subsec:stopping_of_planet_formation}

The formation of planetesimals is stopped when one of the following three conditions is satisfied.
\begin{enumerate}
	\item The pebble density at the iceline is $\rho_{\rm p}/\rho_{\rm g}<1$.
	\item Pebbles from the outermost disk radius reach the iceline, $t= t_{\rm end}$.
	\item The planets are trapped in resonances around $r_{\rm ice}$
\end{enumerate}
When one of the above conditions is satisfied, the simulations are stopped after all formed protoplanets stop growing and migrating.

The time at which Condition 1 is satisfied is estimated based on Equations (\ref{equ:F_pg_nobump}) and (\ref{equ:p_gas_ratio}):
\begin{eqnarray}
	t_{\rho} &=& t_{\rho, {\rm peb}} + t_{\rho,{\rm drift}},
	\label{equ:t_rho}
	\\
	t_{\rho, {\rm peb}} &\simeq&
	%
	1.2\times 10^6\mbox{~yr}
	\left( \frac{M_\star}{0.08M_{\odot}} \right)^{-7/5}
	\left( \frac{R_{\rm disk}}{100\mbox{~au}} \right)^{-3}
	\nonumber \\ &&\times
	\left( \frac{\alpha}{10^{-3}} \right)^{3/2}
	\left( \frac{\tau_{\rm p}}{0.05} \right)^{-3/2}
	\left( \frac{f_{\rm ice}}{19} \right)^{3}
	\left( \frac{\xi}{10} \right)^{-2}
	\nonumber \\ &&\times
	\left( \frac{Z_0}{0.02} \right)^{5}
	\left( \frac{1+ (2t_{\rm drift}) /(3t_{\rm peb) }}{1.5} \right)^{-3},
	\nonumber
\end{eqnarray}
where $t_{\rho,{\rm drift}}$ is $t_{\rm drift}$ at $r_{\rm g}(t_{\rho, {\rm peb}})$.
%
Condition 2 is represented by Equation (\ref{equ:t_end_c}), and Condition 3 is satisfied when the separation between the outermost planet and the iceline inner edge is less than the separation of the 6:5 resonance.
When this condition is satisfied, it is expected that the pileup structure near the iceline is affected by the outermost planet.
This condition can be approximately expressed as
\begin{eqnarray}
	r_N = r_{\rm c} \left( \frac{p+1}{p} \right)^{2(N-1)/3} < r_{\rm ice}.
	\label{equ:r_N}
\end{eqnarray}
The maximum number of planets that satisfies this relationship is $N_{\rm max}$:
\begin{eqnarray}
	N_{\rm max} &\approx& 1+\frac{3}{2} \frac{ \log{(r_{\rm ice}/r_{\rm c})} }{ \log{((p+1)/p)} }
	\nonumber\\
	&=& 1+5.3
	\left( \frac{\log{((p+1)/p)}}{0.3} \right)^{-1}
	\left(
		\frac{ \log{(r_{\rm ice}/r_{\rm c} )} }{1.06}
	\right).
	\nonumber \\
	\label{equ:nmax}
\end{eqnarray}
Condition 3 can be recast to the time condition related to $t_{N}$:
\begin{eqnarray}
	t_{N} = \sum_{i=1}^{N_{\rm max}-1} \left(t_{\rm cross}+t_{\rm pl} \right) + t_{\rm v,ice},
	\label{equ:t_N}
\end{eqnarray}
where $t_{\rm v,ice}$ is the time at which the initial protoplanet is added (Section \ref{subsec:disk}) and is shorter than $t_{\rho}$ and $t_{\rm end}$.
Considering that $N_{\rm max}\sim 6$, we have $t_{N}\gtrsim 5(t_{\rm cross} +t_{\rm pl})+ t_{\rm v,ice}$.
When $t_{\rm end}$ or $t_{\rho}$ is shorter than $t_N$, the number of planets is estimated to be
\begin{eqnarray}
	N \sim \frac{t_{\rm ter} - t_{\rm v,ice}}{t_{\rm cross} +t_{\rm pl}}+1,
	\label{equ:N_tend_trho}
\end{eqnarray}
where $t_{\rm ter} = \min{(t_{\rm end}, t_{\rho})}$ is the termination timescale of the planetesimal formation.
The timescale of $t_{\rm ter} - t_{\rm v,ice}$ represents the duration for all the planetesimals to be able to form progressively at the iceline. 
We refer to this timescale as the planetesimal forming duration.

\begin{figure}
	\plotone{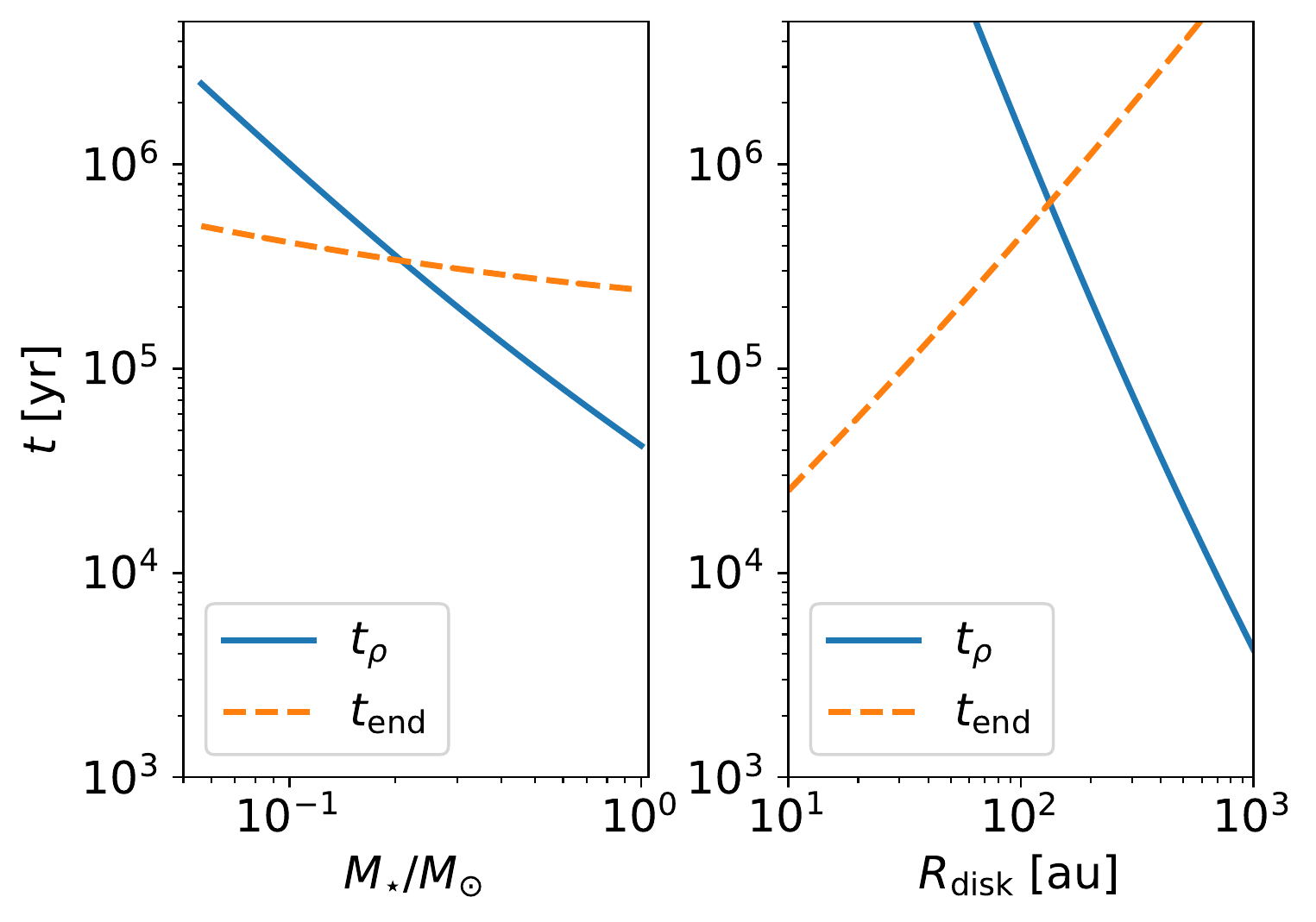}
	\caption{
		Dependences of $t_{\rho}$ (Equation (\ref{equ:t_rho})) and $t_{\rm end}$ (Equation (\ref{equ:t_end_c})) on the stellar mass and disk radius, which
		are expressed as functions of $M_{\star}$ when $R_{\rm disk}=100$~au (left panel) and functions of $R_{\rm disk}$ when $M_{\star}=0.08M_{\odot}$ (right panel).
	}
	\label{fig:Rdisk_Ms_t_term}
\end{figure}

These conditions are shown in Figures \ref{fig:density_evolution} and \ref{fig:Rdisk_Ms_t_term}.
When $M_\star=0.08M_{\odot}$ and $R_{\rm disk}=100$~au, $t_{\rm end}=4.4\times 10^5$~yr is shorter than $t_\rho$.
The stellar mass affects $t_{\rm end}$ and $t_{\rho}$ through $t_{\rm peb}$ and $\mathcal{F}_{\rm p/g}$.
Around a massive star, the planetesimal formation duration is short because both $t_{\rm end}$ and $t_{\rho}$ are short.
The dependences of $t_{\rm end}$ and $t_{\rho}$ on the disk size have opposite outcomes:
$t_{\rm end}$ increases with increasing disk size because the pebble forming time ($t_{\rm peb}$) and the drift time ($t_{\rm drift}$) increase (Equation (\ref{equ:t_end_c}));
by contrast, $t_{\rho}$ decreases with increasing disk size (Equation (\ref{equ:t_rho})).
This dependence arises from $\mathcal{F}_{\rm p/g}\propto \Sigma_{\rm g}\propto R_{\rm disk}^{-1}$; thus, the pebble-to-gas mass flux ratio decreases with increasing disk size because the gas surface density decreases in our model setting.
When $M_\star=0.08M_{\odot}$ and $R_{\rm disk}=200$~au, $t_{\rho}$ is estimated to be $2.2\times 10^5$~yr based on Equation (\ref{equ:t_rho}) and it is $2.8\times 10^5$~yr according to the calculation (Figure \ref{fig:density_evolution}).
Thus, $t_{\rho}$ is shorter than $t_{\rm end}=1.1\times10^6$~yr.

\section{Results} \label{sec:simulation}

\subsection{Time Evolution}\label{subsec:typical_evolution_process}

\begin{figure}
	\plotone{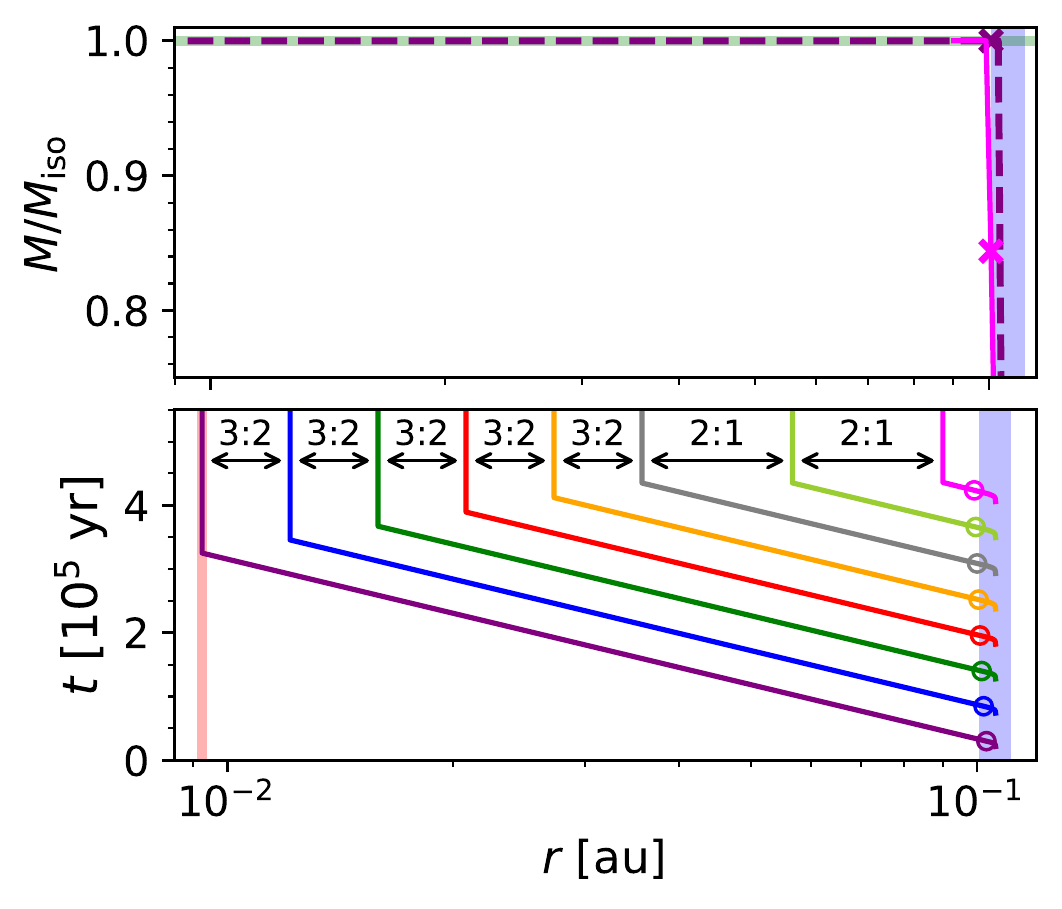}
	\caption{
		Time evolution of protoplanet masses normalized by the isolation mass (top panel) and orbital radii (bottom panel) for $M_\star = 0.08{M_\odot}$, $R_{\rm disk} = 100$~au, and $t_{\rm pl}=4.12\times10^4$~yr.
		The green horizontal line in the top panel represents $M/M_{\rm iso}=1$, the red vertical line in the bottom panel represents $r_{\rm c}$, and the blue vertical lines in both panels represent the iceline region from $r_{\rm ice}(1-\delta_{\rm ice})$ to $r_{\rm ice}(1+\delta_{\rm ice})$.
		The circles in the bottom panel indicate the times at which the protoplanets reach the isolation mass, and the cross symbols in the top panel indicate the times at which protoplanets leave from the iceline region.
		In this case, eight planets are formed, and they are trapped in 3:2, 3:2, 3:2, 3:2, 3:2, 2:1, and 2:1 resonances starting from the innermost pair.
		The top panel presents the evolution of the innermost and outermost planets.
	\label{fig:r_t}}
\end{figure}

In this section, we present the time evolution of protoplanets.
Figure \ref{fig:r_t} shows the time evolution of the protoplanets when $M_\star = 0.08{M_\odot}$, $R_{\rm disk} = 100$~au, and $t_{\rm pl}=4.12\times10^4$~yr.
In this case, eight planets are formed.
The protoplanets grow rapidly around the iceline region and reach the pebble isolation mass at $\sim 0.1$~au.
More specifically, the first four protoplanets reach the pebble isolation mass before they leave from the iceline region due to the high $\rho_{\rm p}/\rho_{\rm g}$ value (Figure \ref{fig:density_evolution}).
The later formed protoplanets have smaller crossing masses because the pebble-to-gas mass flux ratio decreases over time (Equations (\ref{equ:F_pg_nobump}) and (\ref{equ:M_cr_Miso})).
The final protoplanet has the mass ratio $M_{\rm cross}/M_{\rm iso}=0.84$.
Namely, the inner four planets have the same composition as the iceline region, and the outer four planets are mainly composed of 84~\% and more of the iceline materials.

After a protoplanet leaves from the iceline region and $t_{\rm pl}$ is passed, the next protoplanet begins to grow at the iceline.
The typical time interval between the protoplanet appearance is about $5.5\times10^4$ yr, where $t_{\rm pl}$ is the main component of the time interval due to the quick growth of protoplanets (Equations (\ref{equ:t_gr}) and (\ref{equ:t_c_tpl})).
In this case, the protoplanet reaches the isolation mass in the interval, i.e., before the growth of the next protoplanet begins.

The protoplanets migrate inward and become trapped in resonances after the innermost planet reaches $r_{\rm c}$.
The inner five pairs are trapped in 3:2 resonances and the outer two pairs are trapped in 2:1 resonances.
The outer planet pairs are trapped in larger period ratio resonances due to the longer growth timescale.
Because the resonant pair of the planets is given by the closest inner resonance around low-mass stars (Section \ref{sec:migration}), the evolution of their period ratio determines in which resonance planets are trapped.
The period ratios of adjacent planets become the largest value when the outer planet reaches the isolation mass because their migration timescales become identical (Equations (\ref{equ:t_I}) and (\ref{equ:t_II})).
This maximum period ratio is estimated by $(t_{\rm cross}+t_{\rm pl}+ t_{\rm grow})/t_{\rm I}$; i.e., the timescale for the next formed protoplanet reaching the isolation mass is divided by the migration timescale, which is then converted into the typical orbital separation.
The orbital separation given by the time interval between the protoplanet appearance ($t_{\rm cross}+t_{\rm pl}$) is almost the same between planets in individual systems.
The separation between the planets becomes larger as $\mathcal{F}_{\rm p/g}$ decreases, which increases the growth timescale of the next formed protoplanet ($t_{\rm grow}$, Equation (\ref{equ:t_gr})).
Accordingly, the inner planets are trapped in less separated resonances and the outer planets are locked in more separated resonances.

The formation of planetesimals is stopped when the eighth planet is formed.
In this case, Condition 3 is satisfied (Section \ref{subsec:stopping_of_planet_formation}); namely,
the separation between the outermost planets and the innermost radius of the iceline ($r_{\rm ice}(1-\delta_{\rm ice})$) is smaller than the separation of the 6:5 resonance.

\subsection{Resonant Chains of Planets\label{subsec:evolution_result}}

\begin{figure}
	\plotone{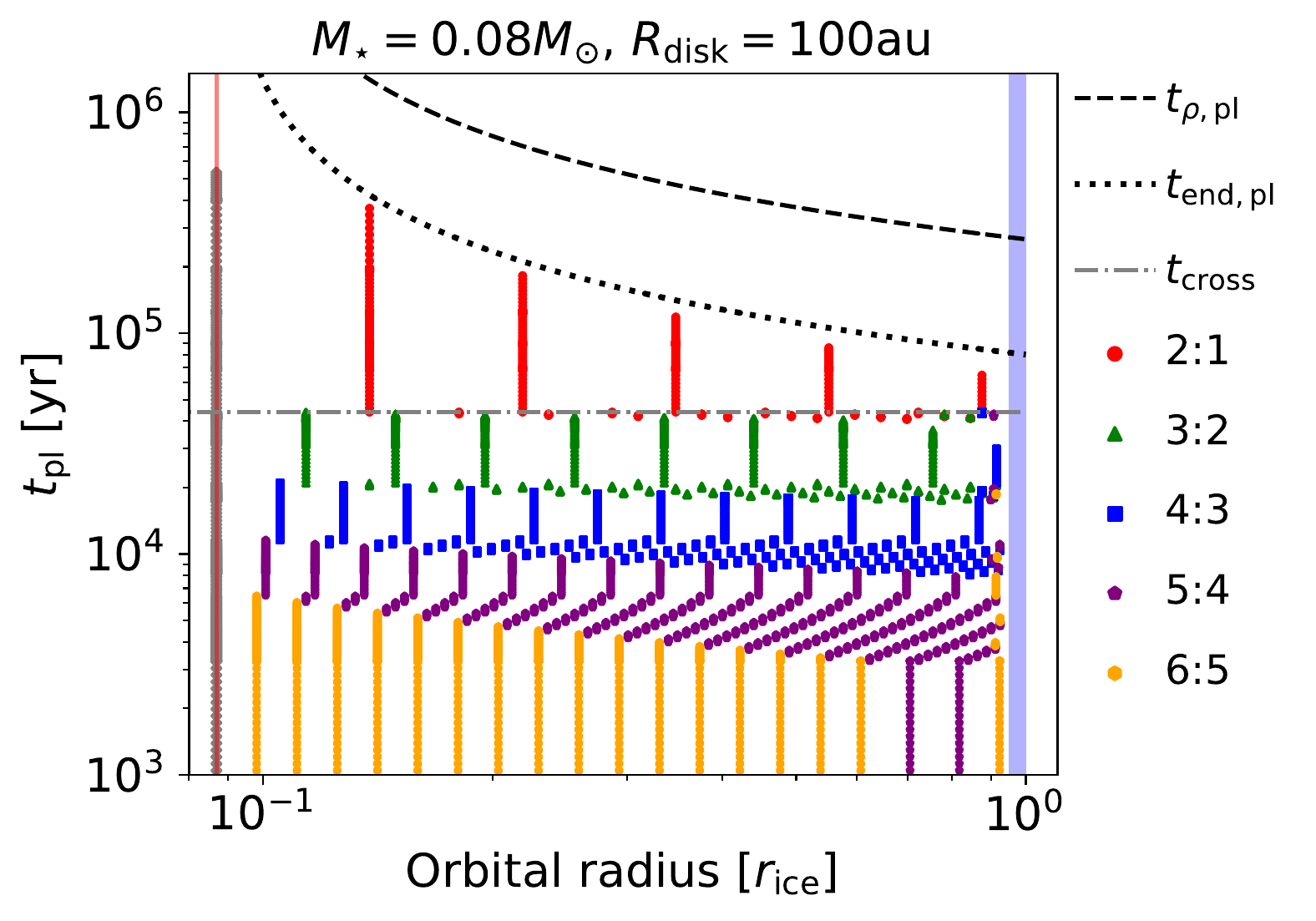}
	\caption{
		Distribution of the orbital radius (in units of $r_{\rm ice}$) of protoplanets is plotted against the protoplanet appearance time ($t_{\rm pl}$) in the protoplanetary systems for $M_\star = 0.08{M_\odot}$ and $R_{\rm disk} = 100~{\rm au}$.
		The innermost planets lie at $r_{\rm c}$, which is indicated by the thin red vertical line.
		The other planets, represented by symbols of various colors, are illustrated based on the resonant relationship of the planet and the inner planet.
		In addition, the dotted line shows
		$t_{\rm end,pl}$ (Equation (\ref{equ:t_rho_end_pl})), the dashed line illustrates
		$t_{\rho,{\rm pl}}$ (Equation (\ref{equ:t_rho_end_pl})), and the dash-dotted line represents
		$t_{\rm cross}$ (Equation (\ref{equ:t_c_tpl})).
	\label{fig:typical_evolution_result}}
\end{figure}

This section presents the simulations with varied $t_{\rm pl}$.
Figure \ref{fig:typical_evolution_result} shows the results for the case with $M_\star = 0.08{M_\odot}$ and $R_{\rm disk} = 100~{\rm au}$.
Evidently, the planets are trapped in more widely separated resonances (i.e., small $p$ resonances) and the number of planets decreases with increasing $t_{\rm pl}$.
Another interesting feature is that the planets in individual systems are trapped in one or two kinds of resonances.
This can be interpreted by the values of $(t_{\rm cross}+t_{\rm pl}+ t_{\rm grow})/t_{\rm I}$:
While the outer planets have longer $t_{\rm grow}$, planets have similar values in individual systems, which means that the maximum separations of planets are similar.
It is worth noting that the outermost planet pairs are often trapped in a closely separated resonance than the inner ones.
This is because the separation between the second outermost planet and the iceline inner edge is small.
In addition, the resonant trapping is considered after the protoplanets leave from the iceline region.

In Figure~\ref{fig:typical_evolution_result}, $t_{\rm cross}$, $t_{\rho,{\rm pl}}$, and $t_{\rm end,pl}$ are also plotted, where
\begin{eqnarray}
	t_{\rho,{\rm pl}} = \frac{t_{\rho} - t_{\rm v,ice} }{ 1.5 \log_2{(r/r_{\rm c})} },\
	t_{\rm end,pl} = \frac{t_{\rm end} - t_{\rm v,ice} }{ 1.5 \log_2{(r/r_{\rm c})} }.
	\label{equ:t_rho_end_pl}
\end{eqnarray}
These timescales are derived from Equations (\ref{equ:N_tend_trho}) and (\ref{equ:r_N}) with $p=1$.
Using these timescales, the termination Conditions 1 or 2 can be written as ($t_{\rm pl}$ or $t_{\rm cross}$) $\gtrsim$ ($t_{\rho,{\rm pl}}$ or $t_{\rm end,pl}$), and we can present these conditions in the figure.
For $M_\star = 0.08{M_\odot}$ and $R_{\rm disk} = 100~{\rm au}$, the curve corresponding to $t_{\rm end,pl}$ agrees with the orbital distribution of planets when $t_{\rm pl} \gtrsim10^5$~yr.

When $t_{\rm pl}< 10^5$~yr, the planets fill orbital radii between $r_{\rm c}$ and $r_{\rm ice}$.
The number of planets is well represented by Equation (\ref{equ:nmax}).
When $t_{\rm pl}< 2\times 10^4$~yr, more than 10 planets are trapped in 4:3 or closer resonances.
These planets could cause orbital instability and experience giant impacts after the disk gas removal \citep{matsumoto2012orbital, Matsumoto&Ogihara2020}.
The resulting giant impacts continue to grow the planet mass up to the so-called ejection mass and even eject planets, leading to non-resonant planetary systems and free-floating planets due to the large total mass of planets \citep{Matsumoto2020}.
This suggests that observed planets in resonant chains would be mainly composed of 3:2 or 2:1 resonances.
In the following sections, we adopt $N_{\rm crit}=10$ as the criterion for the stable planet systems in resonant chains and discuss the final configuration of planets.

The abundance of the iceline materials in the formed planets is almost as high as that in the case that $t_{\rm pl}=4.12\times 10^4$~yr.
This is because planets are formed in the high $\mathcal{F}_{\rm p/g}$ environment when $M_\star = 0.08{M_\odot}$ and $R_{\rm disk} = 100~{\rm au}$.
In this stellar mass and disk size, the termination Condition 1, $\rho_{\rm p}/\rho_{\rm g}<1$, is not satisfied (Figure \ref{fig:typical_evolution_result}).
All planets are formed in the environment of the high pebble-to-gas density ratio at the iceline (Figure \ref{fig:density_evolution}), i.e., high $\mathcal{F}_{\rm p/g}$ (Equation (\ref{equ:p_gas_ratio})).
The formed planets are mainly composed of the iceline material since planets grow quickly in the high $\mathcal{F}_{\rm p/g}$ cases (Equation (\ref{equ:M_cr_Miso})).

\subsection{Disk Size Dependence \label{subsec:disk_size_dependence}}

\begin{figure}
	\plotone{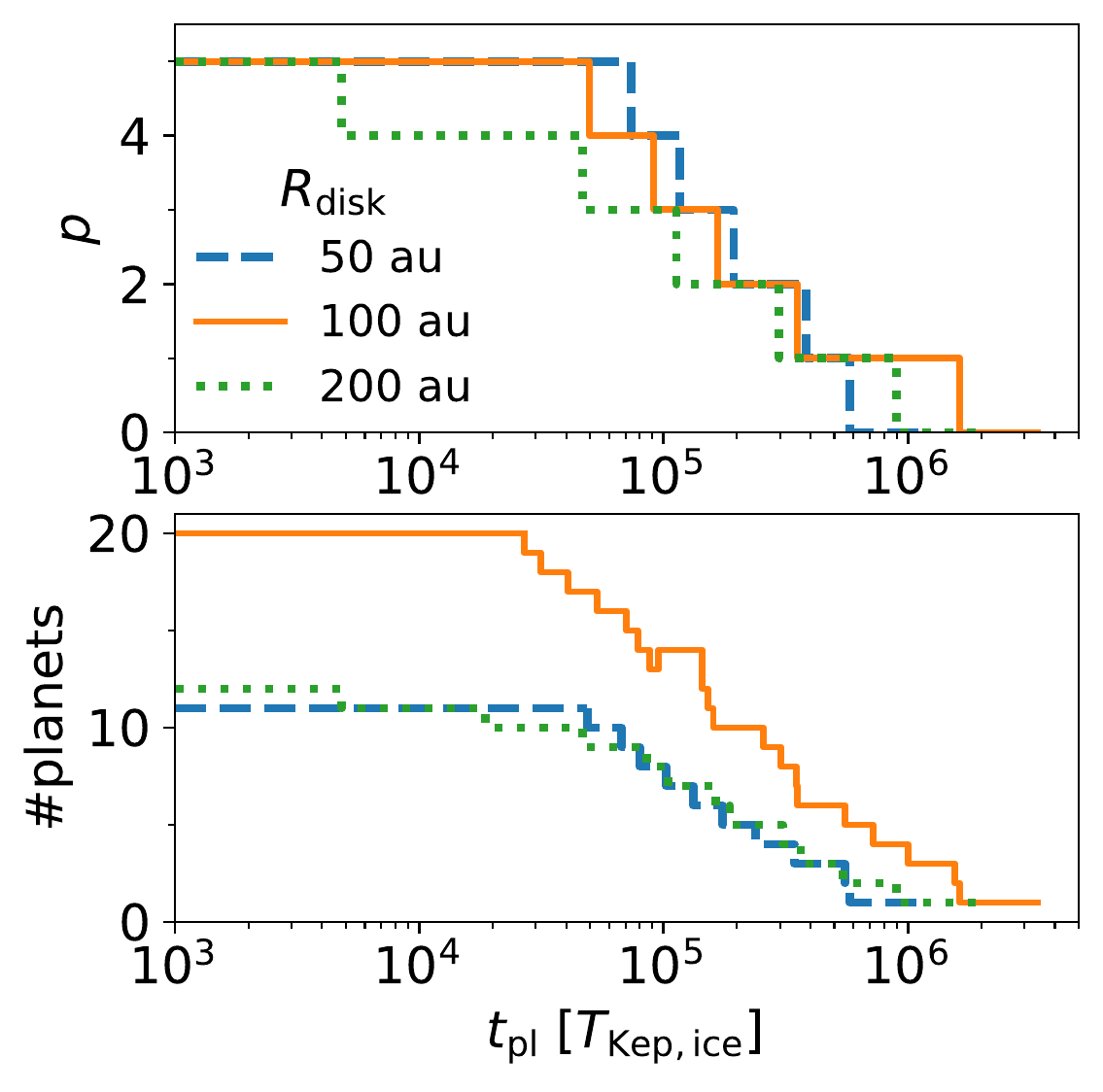}
	\caption{
	Resonance of the innermost planet pairs ($p+1:p$, the top panel) and the number of planets (the bottom panel) are plotted as functions of the protoplanet appearance time normalized by the Kepler time at the iceline ($t_{\rm pl}~[T_{\rm Kep, ice}]$) for different disk sizes when $M_{\star}=0.08M_{\odot}$.
	\label{fig:tpl_np_Rdisk}}
\end{figure}

The disk size affects the pebble-to-gas mass flux ratio ($\mathcal{F}_{\rm p/g}\propto R_{\rm disk}^{-1}$, Equation (\ref{equ:F_pg_nobump})) and the termination condition of the protoplanet formation.
Figure \ref{fig:tpl_np_Rdisk} shows the number of planets and the resonance in which the innermost planet pairs are trapped.
These indicate the configurations of the final planets because the planets are trapped in resonant chains composed of one or two resonances (Section \ref{subsec:evolution_result}).
In Figure~\ref{fig:tpl_np_Rdisk}, we show the results for the $R_{\rm disk}=50$ au, 100 au, and 200 au cases around $0.08M_{\star}$ stars.
Interestingly, the relationships between the innermost resonance and $t_{\rm pl}$ in these results are similar.
The planets tend to be trapped in more widely separated resonances (i.e., smaller $p$ resonances) as the disk size increases.
This is because $\mathcal{F}_{\rm p/g}$ decreases and $t_{\rm grow}$ increases with increasing disk size.

The number of planets in the $R_{\rm disk}=100$~au case is the largest for any $t_{\rm pl}$.
This is because the termination timescale of the planetesimal formation ($t_{\rm ter}=\min{(t_{\rho}, t_{\rm end})}$) is the longest in the $R_{\rm disk}=100$~au case for these three cases (Figure \ref{fig:Rdisk_Ms_t_term}).
The termination timescales are $1.8\times 10^5$~yr in the $R_{\rm disk}=50$~au case, $4.4\times 10^5$~yr in the $R_{\rm disk}=100$~au case, and $2.2\times 10^5$~yr in the $R_{\rm disk}=200$~au case.
While the planets fill orbital radii in the $R_{\rm disk}=100$~au case when $t_{\rm pl}< 10^5$~yr, the numbers of planets are regulated by $t_{\rm end}$ in the $R_{\rm disk}=50$~au case and by $t_{\rho}$ in the $R_{\rm disk}=200$~au case.
As a result, the numbers of planets of the $R_{\rm disk}=50$~au and $R_{\rm disk}=200$~au cases are similar and are smaller than in the $R_{\rm disk}=100$~au case.

The maximum disk size for the formation of planets in resonant chains in this scenario is determined based on $t_{\rm ter}(=t_{\rho})< t_{\rm cross}$.
When this condition is satisfied, only one planet is formed and the next planetesimal cannot be formed.
The maximum disk size is $R_{\rm disk} \simeq 320$~au around $0.08M_{\odot}$ stars.
Furthermore, no planets are formed when $R_{\rm disk} \geq 360$~au because $\rho_{\rm p}/\rho_{\rm g}$ is less than 1 even at $t=t_{\rm v,ice}$.

The disk size can affect the resonant chains.
In the $R_{\rm disk}=100$~au case, after the gaseous disk is dissipated, the planets can escape from resonant orbits due to orbital instabilities when they are trapped in 4:3 or more closely separated resonances because the number of planets exceeds $N_{\rm crit}$.
However, in the $R_{\rm disk}=50$~au and $R_{\rm disk}=200$~au cases, planets trapped in 4:3 or more closely separated resonances do not always cause orbital instabilities because the number of planets is less than $N_{\rm crit}$.
If any exoplanetary systems composed of Earth-mass planets trapped in 4:3 resonances are observed around $0.08M_{\odot}$ stars, they would have formed in disks with sizes of $R_{\rm disk}\sim50$~au or $\sim 200~$au.

The compositions of the formed planets depend on the disk size since $\mathcal{F}_{\rm p/g}\propto R_{\rm disk}^{-1}$.
This indicates that while the number of planets and their resonances are similar between the $R_{\rm disk}=50$~au and 200~au cases, their compositions are different.
In the $R_{\rm disk}=50$~au cases, planets are composed of the iceline materials.
In contrast, in the $R_{\rm disk}=200$~au cases, the mass fractions of the iceline materials in planets are between 0.77 and 0.49.

\subsection{Stellar Mass Dependence \label{subsec:stellar_mass_dependence}}

\begin{figure}
	\plotone{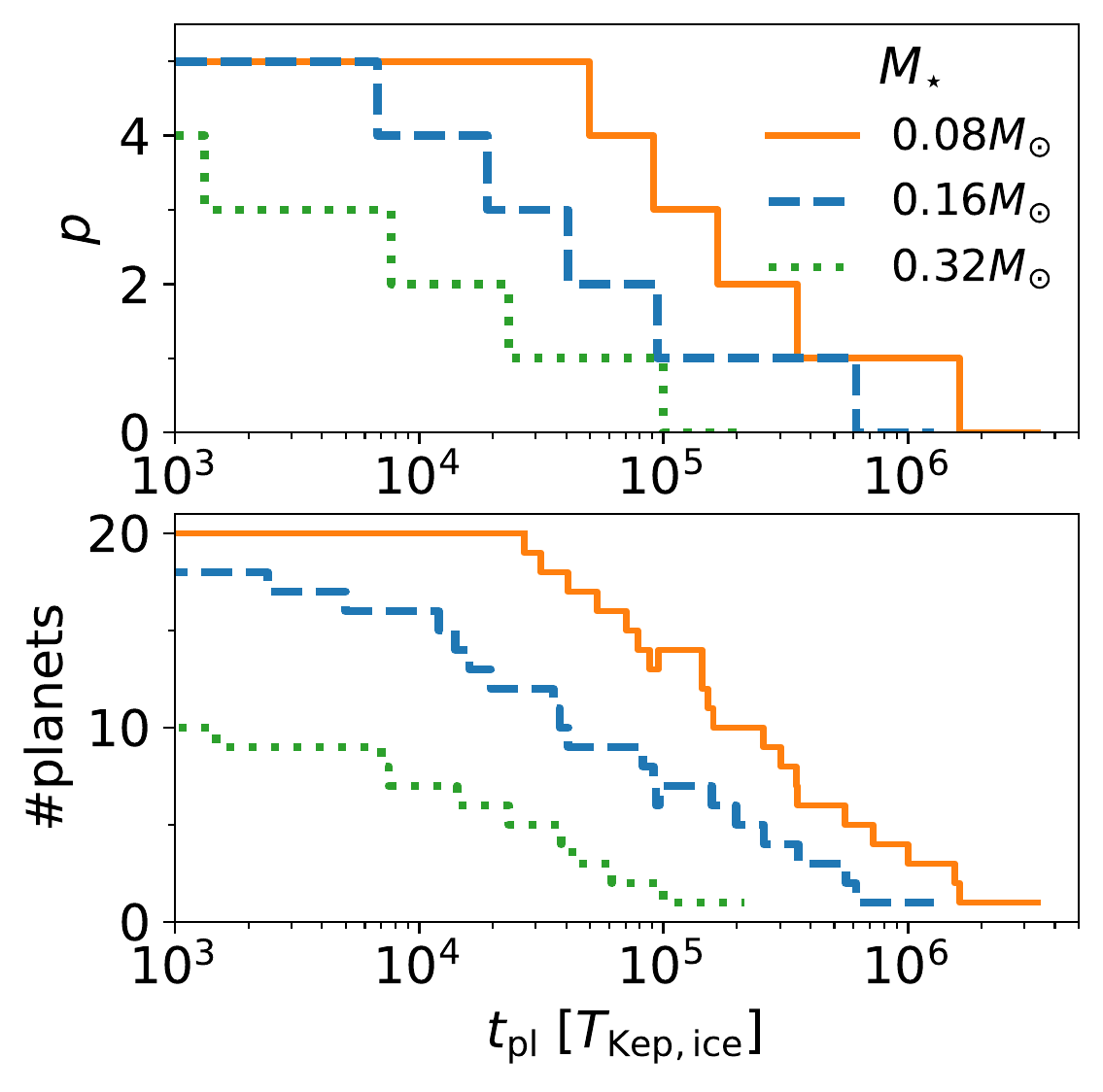}
	\caption{
		Similar to Figure \ref{fig:tpl_np_Rdisk}; however, the figure presents the protoplanet appearance time normalized by the Kepler time at the iceline ($t_{\rm pl}~[T_{\rm Kep, ice}]$) for different stellar masses when $R_{\rm disk}=100$~au.
		}
	\label{fig:tpl_np_Ms}
\end{figure}

The stellar mass dependences on the resonance of the innermost pairs and the number of planets are shown in Figure \ref{fig:tpl_np_Ms}.
As the stellar mass increases, the innermost planets become trapped in wider separation resonances (small $p$).
The stellar mass dependence on the period ratio can be estimated using $(t_{\rm cross}+t_{\rm pl}+ t_{\rm grow})/t_{\rm I}$.
We note that the stellar mass dependence is included in the pebble-to-gas mass flux ratio, $\mathcal{F}_{\rm p/g}\propto M_\star^{-7/15}$ in addition to the insignificant contribution from $t_{\rm drift}/t_{\rm peb}$.
The stellar mass dependence on $(t_{\rm cross}+t_{\rm pl}+ t_{\rm grow})/t_{\rm I}$ is positive; i.e., the period ratio increases with increasing stellar mass.

Fewer planets are formed around more massive stars.
This is because the planetesimal forming duration becomes shorter.
Around massive stars, the termination timescale of the planetesimal formation becomes shorter (Figure \ref{fig:Rdisk_Ms_t_term}) and the viscous timescale at the iceline becomes longer (Equation (\ref{equ:viscous_time_ice})).
Because the planetesimal formation takes place from $t_{\rm v, ice}$ to $t_{\rm ter}$, the number of planets decreases with increasing stellar mass.
In our results, no planets are formed when $M_{\star}\geq 0.4M_{\odot}$ because $\rho_{\rm p}/\rho_{\rm g}$ is less than 1 at $t_{\rm v, ice}$.
This is one possible explanation of why the fraction of resonant or near resonant planetary systems around low-mass stars seems to be high.

Next, we consider the condition in which $N\leq N_{\rm crit}$ planets are formed.
The protoplanet appearance time that satisfies $N\leq N_{\rm crit}$ is longer around less massive stars.
This condition is satisfied for the following cases:
$1.6\times10^5 \leq t_{\rm pl}/T_{\rm Kep, ice} \leq 1.6\times10^6 $ in the $M_\star=0.08M_{\odot}$ case,
$3.8\times10^4 \leq t_{\rm pl}/T_{\rm Kep, ice} \leq 6.2\times10^5$ in the $M_\star=0.16M_{\odot}$ case,
and $8.1\times10^2 \leq t_{\rm pl}/T_{\rm Kep, ice} \leq 1.0\times10^5 $ in the $M_\star=0.32M_{\odot}$ case.
The protoplanet appearance time to form the stable planet systems in resonant chains varies with the stellar mass. 
We suggest that the tyical protoplanet appearance time can be estimated based on the fraction of planets near resonant chains as the function of the stellar mass.

The stellar mass affects the compositions of the formed planets due to the stellar mass dependence of $\mathcal{F}_{\rm p/g}$.
When we consider the planet formation around a more massive star, $\mathcal{F}_{\rm p/g}$ becomes smaller, and planets are more silicate-rich.
In the case that $M_\star=0.32M_{\odot}$, the mass fractions of the iceline materials in planets are between 0.56 and 0.49.

\section{Discussion} \label{sec:discussion}

In this paper, we present a time evolution model for multiple-planet formations based on the pebble-driven planet formation scenario \citep{ormel2017formation}.
Although the relationship between $t_{\rm pl}$ and the resonant chains of the planets can be determined, our results are based on a number of assumptions.
In the following subsections, the assumptions and the formation of the TRAPPIST-1 system are discussed.

\subsection{Disk Structure} \label{subsec:Disk_Structure}

In the results of this study, the number of planets depends on the termination time of the planetesimal formation ($t_{\rm ter}$), which is the minimum of the final pebble reach time ($t_{\rm end}$) and the time when $\rho_{\rm p}/\rho_{\rm g}<1$ ($t_{\rho}$).
We assume that the pebbles are formed through the coagulation of icy grains that originally exist in the outer disk.
Namely, we neglected the dust supply from the remnant star-forming core and surrounding interstellar medium, while these would be important to explain the discrepancy between the observed dust mass and observed exoplanet mass \citep{Manara+2018}.
These processes affect our results when the following conditions are satisfied:
they provide sufficient pebble mass fluxes for keeping $\rho_{\rm p}/\rho_{\rm g}>1$;
the durations of these processes are longer than the termination timescale of the planetesimal formation ($t_{\rm ter}< 10^6$~yr).
Such a high and long-term infall onto the star-disk system was suggested by the numerical simulation of turbulent molecular clouds \citep{Padoan+2014}.
While it is not clear whether this mass discrepancy is due to the dust supply or dust scattering \citep{Ueda+2020}, a high and long-term infall would help to grow more planets.

We considered the evolution of protoplanets in the early phase of the disk evolution ($t_{\rm ter}<10^6$~yr).
This short timescale is the reason why the evolution of the gas disk is not taken into account.
The lifetime of gas disks around low-mass stars is typically $\gtrsim 3$~Myr \citep{Luhman2012}.
The gas accretion does not change significantly before $<1$~Myr.
Consequently, the location of the iceline does not move since the viscous heating, which arose from the gas accretion, is almost unchanged \citep{oka2011evolution}.
Moreover, \cite{Takahashi_SZ&Muto2018} showed that the disk evolution can be influenced by the MHD disk wind \citep[e.g.,][]{Suzuki&Inutsuka2009, Bai2014}, which enables the creation of a ring--hole structures within $10^6$~yr.
Such disk evolution can also affect the inward migration of planets \citep{Ogihara+2015}.
The efficiency of the disk evolution due to the MHD disk wind around M~dwarfs is a key parameter that should be considered in the final planet configuration, i.e., to determine whether the planets are in resonant chains.

We adopt $f_{\rm ice}=19$ in all simulations.
If we consider smaller $f_{\rm ice}$, our results change as follows:
termination Condition 1 is quickly satisfied since $t_{\rho, {\rm peb}}\propto f_{\rm ice}^3$ (Equation (\ref{equ:t_rho}));
planets are formed in narrower parameter spaces of $M_\star$ and $R_{\rm disk}$ due to short $t_{\rho, {\rm peb}}$ (Figure \ref{fig:Rdisk_Ms_t_term});
the number of the planets in resonances becomes smaller due to short $t_{\rm ter}$ and long $t_{\rm grow}$ (Equation (\ref{equ:t_gr}));
planets are more silicate-rich (Equation (\ref{equ:M_cr_Miso})).

\subsection{Pebble Accretion}\label{subsec:discuss_pebble}

\subsubsection{Isolation Mass}\label{subsec:Miso}

We adopt a simple formula as the pebble isolation mass (Equation (\ref{equ:Miso})).
According to recent hydrodynamical simulations, which provided more accurate expressions \citep[e.g.,][]{Bitsch+2018, Ataiee+2018}, the pebble isolation mass is about 59\% of our expression \citep[here, we use the expression of][]{Ataiee+2018}.
This less massive isolation mass affects our results, slightly.
When the isolation mass is less massive than our expression, the growth timescale becomes shorter.
Since the resonances in which planets are trapped depend on the timescale over which the next protoplanet reaches the isolation mass (Section \ref{subsec:typical_evolution_process}), the relation between the resonances and $t_{\rm pl}$ is slightly changed.
For example, in the parameter set of Figure \ref{fig:r_t}, eight planets are formed and all are trapped 3:2 resonances when we consider the isolation mass given by \cite{Ataiee+2018}.
This also affects compositions of planets: all planets reach the isolation mass around the iceline region in this parameter set.

\subsubsection{Filtering}\label{subsec:filter}

We do not consider the pebble filtering effect, while we consider the formation of multiple-planet systems.
In our model, protoplanets quickly grow up around the iceline region (Equations (\ref{equ:t_gr}) and (\ref{equ:M_cr_Miso})).
Earlier formed protoplanets obtain the isolation mass before the next protoplanets begin to grow since they satisfy the condition that the growth time to the isolation mass is shorter than $t_{\rm cross}+t_{\rm pl}$.
This condition is not satisfied in the case of the later formed protoplanets in large $R_{\rm disk}$ disks or around massive stars.
These protoplanets accrete filtered pebbles.
Filtering by a single protoplanet is inefficient (Equation (\ref{equ:eps_2}), see also \cite{Lambrechts&Johansen2014}).
The efficiency of filtering would depend on the size distribution of planetesimals around the iceline \citep{Guillot+2014}.

\subsection{Appearance of Protoplanets} \label{subsec:t_pl}

We consider that protoplanet precursors of 100~km in size successively appear, separated by a time $t_{\rm pl}$ after the preceding protoplanet has crossed the iceline inner edge. 
In this section, we disucss $t_{\rm pl}$, although we need the hydrodynamical simulations that include planetesimals and pebbles around the iceline to consider a more realistic picture.
The appearance timescale of the protoplanet would be given by the planetesimal formation timescale and its growth timescale.
The planetesimal formation timescale via the streaming instability depends on the initial local pebbles-to-gas ratio, the Stokes numbers of pebbles, and the strength of turbulence \citep[e.g.,][]{Youdin&Johansen2007, CL2020,Umurhan2020}.
Some simulations show that the quick growth of clumps or filaments operate on timescales of $\lesssim 10^3T_{\rm Kep}$\citep[e.g.,][]{johansen2009particle,yang2017concentrating}.
However, the turbulent diffusion prolongs the growth timescale of the streaming instability.

The initial growth of planetesimals that are precursors of protoplanets around the iceline is pebble accretion of scattered planetesimals and/or runaway growth of planetesimals \citep{Liu_B+2019, Schoonenberg+2019}.
These scattering and runaway growth timescales can be estimated by the dynamics of planetesimals  \citep[e.g.,][]{Kokubo&Ida2000,Liu_B+2019}.
The timescale of the viscous stirring, which provides the timescale for planetesimals to scatter, is $\sim 10^5T_{\rm Kep}$.
The timescale of the runaway growth is $\sim 10^4T_{\rm Kep}$ -- $10^5T_{\rm Kep}$.
These timescales agree with the protoplanet appearance time in \cite{Schoonenberg+2019}.
These suggest that $t_{\rm pl}$ would be $\sim 10^4T_{\rm Kep,ice}$ -- $10^5T_{\rm Kep,ice}$.
It would be worth noting that these timescales, especially for the runaway growth, depend on the surface density of planetesimals.
The above timescales would change if the surface density of planetesimals changes significantly.
The planetesimal forming efficiency is important to consider the subsequent growth of formed planetesimals.

\subsection{Application to TRAPPIST-1} \label{subsec:Trappist1}

TRAPPIST-1 has a mass of $\approx0.08M_{\odot}$ and hosts seven transiting Earth-mass planets with period ratios of 8:5, 5:3, 3:2, 3:2, 4:3, and 3:2 starting from the innermost pair \citep{gillon2017seven,luger2017seven}.
It was suggested that the inner three planets are in each 3:2 resonant chain and that they experience the expansions of orbital separations induced by the stellar tide \citep{Papaloizou+2018} or by the magnetospheric rebound mechanism \citep{ormel2017formation, Liu_B+2017}.
Based on these studies, we consider the planets in the initial resonant chain composed of 3:2 and 4:3 resonances.
In our results, seven planets are trapped in 3:2 resonant chains in the $R_{\rm disk}=200$~au case (Figure \ref{fig:tpl_np_Rdisk}).
In this case, the particular protoplanet appearance time at which the planets are trapped in the 3:2 resonances and 4:3 resonances is about $10^5T_{\rm Kep,ice}$.
When $t_{\rm pl}$ of the TRAPPIST-1 g is slightly shorter than $10^5T_{\rm Kep,ice}$ and the others are slightly longer, our results reproduce the planets trapped in the 3:2, 3:2, 3:2, 3:2, 4:3 and 3:2 resonant chains.

Next, we discuss the compositions of TRAPPIST-1 planets.
The studies of interior modeling suggested that the water mass fractions of TRAPPIST-1 planets have the following two features: their water mass fractions are $\lesssim 25$~wt\%; these fractions are uniform or increasing with orbital periods \citep[e.g.,][]{Dorn+2018, Unterborn+2018,Agol+2020}.
Our model predicts that planets are mainly composed of iceline material.
The mass fractions of the iceline material contents in the planets are between 0.77 (the innermost) and 0.50 (the outermost).
While our results suggest seemingly higher in the water mass fractions, the water mass fractions of TRAPPIST-1 planets can be explained if we consider the accretion of silicate grains released by the sublimation of icy pebbles around the iceline \citep{hyodo2019formation}.
It would be worth noting that \cite{Agol+2020} showed that the water mass fractions of TRAPPIST-1 planets are about equal or less than 5~wt\% if they have Earth-like interior structures. 
Such small fractions of waters are explained only when almost all iceline materials are silicate. 
In our results, the outer planets have less iceline materials.
To explain the dependence of the water mass fractions of TRAPPIST-1 planets on orbital periods, it is needed that the iceline materials become more water-rich as icy pebbles formed at the outer region reach.

\section{Conclusion} \label{sec:conclusion}

Recent observations have revealed the existence of multiple-planet systems around low-mass stars, which comprises planets with masses of $\sim 1M_{\oplus}$ planets.
Interestingly, planets are near resonant orbits in most of these systems.
In this study, we considered the formation of multiple-planet systems around low-mass stars.
We construct a time evolution model of protoplanets around low-mass stars based on the scenario proposed by \cite{ormel2017formation}.
The model considers the formation of planetesimals at the iceline and their growth due to pebble accretion.

We find that the protoplanet appearance timescale ($t_{\rm pl}$) plays an important role in the configuration of resonant trapping and in the number of planets.
When $t_{\rm pl}$ is short, many planets are formed, and they are trapped in more closely separated resonances.
These planets cause orbital instabilities after the disk gas removal, and they are expected to end up being planets in non-resonant orbits and even free-floating planets.
As $t_{\rm pl}$ increases, the number of planets decreases, and the planets become trapped in more widely separated resonances.
This indicates that there exists a range of $t_{\rm pl}$ for forming planets in resonant chains.
The formed planets are trapped in the resonant chains that are composed of one or two kinds of resonances.
We predict that inner planets have richer iceline materials.
It is worth noting that our simulation results are based on simple analytical estimations.
Nevertheless, these estimations provide a way for expanding the formation of the single-planet system to the formation of the multiple-planet system.

The disk size and the stellar mass are used as parameters in simulations.
We find that the number of planets varies with the disk size because the planetesimal forming duration, which represents the duration determined by the condition that the pebble density is larger than the gas density at the iceline, changes.
The disk size also affects the compositions of planets.
We can predict the disk size and $t_{\rm pl}$ of observed planets from their features such as the resonances, number of planets, and their compositions.
Our model can well reproduce the TRAPPIST-1 system when the disk size is 200~au and $t_{\rm pl}$ is about $10^5$ orbital periods around the iceline.
We also find that the range of $t_{\rm pl}$ for forming planets in resonant chains depends on the stellar mass. 
Resonant planets are not formed when the stellar mass is larger than $0.4M_{\odot}$.
This is a possible explanation of why there are not many planets near resonant orbits around stars with masses of $\sim 1M_{\odot}$.

\acknowledgments

We thank Min-Kai Lin for useful discussions.
We thank the referee for constructive comments.
This research is supported by MOST in Taiwan (grant 105-2119-M-001-043-MY3 and 109-2112-M-001-052-) and the ASIAA Summer Student Program.
Numerical analyses were in part carried out on analysis servers at Center for Computational Astrophysics, National Astronomical Observatory of Japan.

\bibliography{LMG}{}
\bibliographystyle{aasjournal}

\end{document}